\documentclass[conference]{IEEEtran}
\IEEEoverridecommandlockouts

\usepackage{cite}
\usepackage{amsmath,amssymb,amsfonts}
\usepackage{algorithmic}
\usepackage{graphicx}
\usepackage{textcomp}
\usepackage{xcolor}
\usepackage{tabularray}

\usepackage{background}
\usepackage{xcolor}
\usepackage{hyperref}

\hypersetup{
  pdfborder={0 0 0}, 
}

\def\BibTeX{{\rm B\kern-.05em{\sc i\kern-.025em b}\kern-.08em
    T\kern-.1667em\lower.7ex\hbox{E}\kern-.125emX}}

\backgroundsetup{
  scale=1,
  color=black,
  opacity=1,
  angle=0,
  position=current page.south,
  vshift=10pt,
  contents={\textcolor{red}{© 2023 IEEE. For personal use only. Published version under doi:  \href{https://doi.org/10.1109/MASS58611.2023.00082}{10.1109/MASS58611.2023.00082}. }}
}

\begin{document}

\title{Enabling Architecture for Distributed Intelligent Network Softwarization for the Internet of Things}

\author{\IEEEauthorblockN{Mohamed Ali Zormati}
\IEEEauthorblockA{\textit{Heudiasyc UMR 7253} \\
\textit{Université de Technologie de Compiègne}\\
Compiègne, France \\
zormati@ieee.org}
\and
\IEEEauthorblockN{Hicham Lakhlef}
\IEEEauthorblockA{\textit{Heudiasyc UMR 7253} \\
\textit{Université de Technologie de Compiègne}\\
Compiègne, France \\
hicham.lakhlef@hds.utc.fr}
}

\maketitle

\begin{abstract}
The Internet of Things (IoT) is becoming a part of everyday life through its various sensing devices that collect valuable information. The huge number of interconnected heterogeneous IoT devices poses immense challenges, and network softwarization techniques are an adequate solution to these concerns. Software Defined Networking (SDN) and Network Function Virtualization (NFV) are two key softwarization techniques that enable the realization of efficient, agile IoT networks, especially when combined with Machine Learning (ML), mainly Federated Learning (FL). Unfortunately, existing solutions do not take advantage of such a combination to strengthen IoT networks in terms of efficiency and scalability. In this paper, we propose a novel architecture to achieve distributed intelligent network softwarization for IoT, in which SDN, NFV, and ML combine forces to enhance IoT constrained networks.
\end{abstract}


\section{Introduction}
The Internet of Things (IoT) connects billions of heterogeneous, constrained devices to each other and to the Internet \cite{b1}. By collecting valuable information, IoT sensing objects impact all areas of human life (e.g., healthcare, transportation, logistics, etc.). The massive number of devices generating an immense amount of data makes IoT networks difficult to manage and control. Thus, IoT networks face several challenges that need to be addressed (e.g., scalability, energy efficiency, security, etc.) \cite{b3}.

When combined with IoT, network softwarization addresses network concerns while providing some benefits (e.g., programmability) \cite{b4}. It mainly consists of Software Defined Networking (SDN) and Network Function Virtualization (NFV) techniques, which enable easier management, configuration, and scalability of the network. SDN creates a programmable network by decoupling the control plane (i.e., the control logic) from the data plane (i.e., the network elements) \cite{b5}. NFV decouples network functions (e.g., load balancing, firewall) from physical network devices and runs them on general-purpose hardware via virtualization technologies \cite{b3}.

Recently, Machine Learning (ML) has become a promising approach to bring intelligence to the network when combined with network softwarization techniques, as it is an enabling technology to make networks self-aware, self-adaptive, and self-managed \cite{b6}. Federated Learning (FL) is an attractive technique to overcome resource constraints in IoT networks and enable efficient machine learning while improving the data privacy, as raw data is not shared between devices\cite{b7}. Therefore, the combination of SDN, NFV, and  ML is a key solution to alleviate the constraints of IoT networks.

Although many research efforts have been made to investigate these three technologies and their combination, existing solutions generally consider SDN and omit NFV \cite{b8} \cite{b9}, even though they are mutually reinforcing. We also note the lack of smart softwarization solutions for IoT networks \cite{b1} \cite{b4}. To the best of our knowledge, no work has been done to propose a distributed intelligent network softwarization architecture for IoT networks, based on SDN, NFV, and FL, thus boosting network efficiency and scalability. Our current work aims to fill this gap.

In the remainder of this paper, we present our architecture proposal in detail in Section II. In Section III, we conclude the paper and present the main future directions.

\section{Overview of the Solution}
Here, we present an enabling architecture to achieve distributed intelligent network softwarization for IoT. Considering the conventional IoT architecture, which mainly consists of sensors and controllers, and to mitigate IoT network issues (e.g., heterogeneity, scalability), we add SDN and NFV components as these are the main network softwarization techniques to obtain an IoT network softwarization architecture. As the number of IoT devices is growing exponentially, having a single SDN controller becomes an issue in terms of reliability. Therefore, we opt for a distributed hierarchical SDN controller design where multiple controllers coexist and are coordinated by a root controller. Such a design allows the application of distributed learning approaches, namely Federated Learning (FL). It will allow the system not only to take advantage of ML techniques (especially Deep Reinforcement Learning) for autonomous efficient decision making, but also to build a global knowledge without having to share raw data between the controllers, as each controller acts as a learning node, and where the root controller acts as an FL aggregator.

\begin{figure*}[htbp]
\centerline{\includegraphics[scale=0.37]{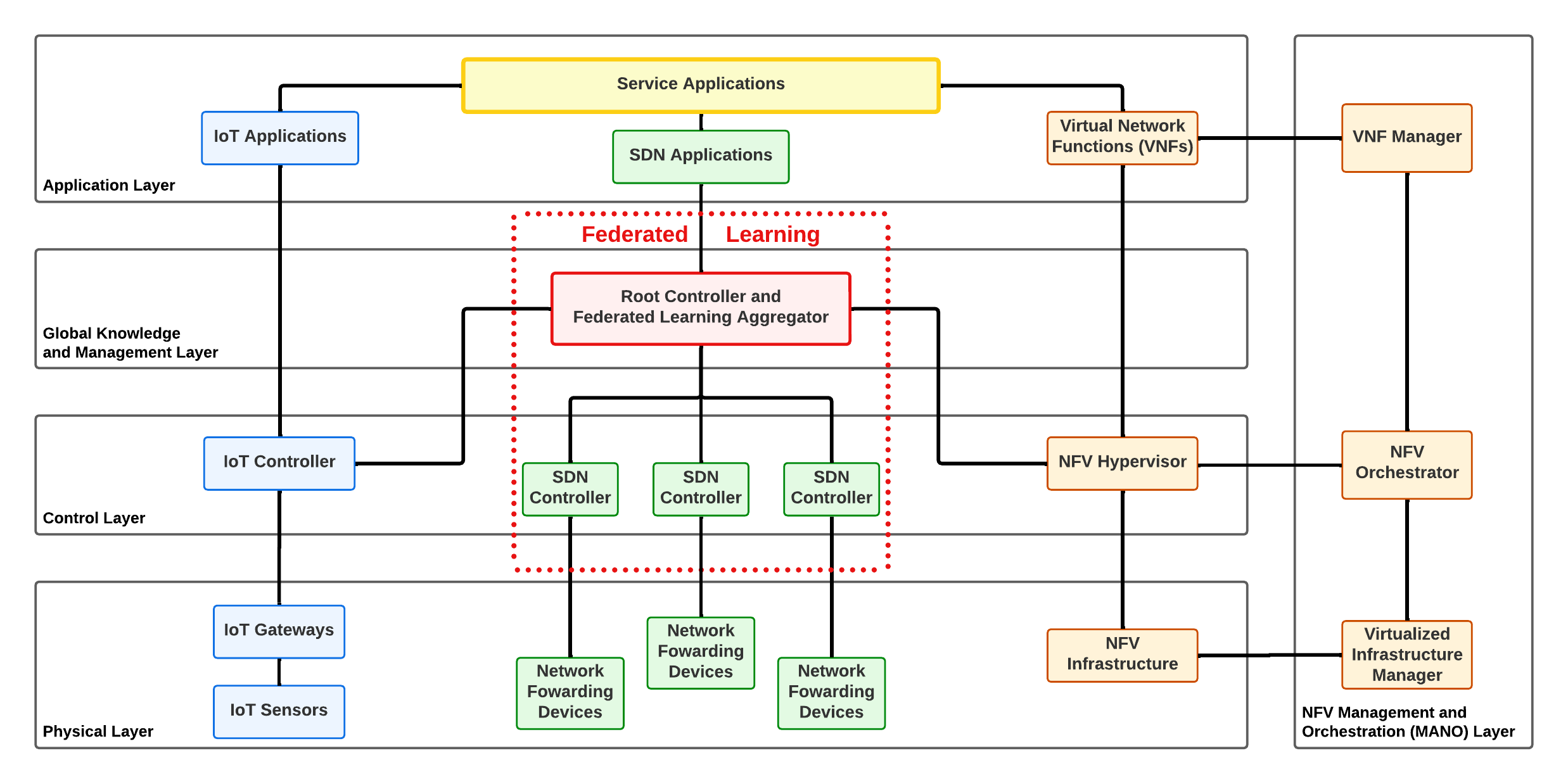}}
\caption{Distributed Intelligent Network Softwarization Architecture for IoT}
\label{fig}
\end{figure*}

Figure 1 shows the proposed architecture, which consists of five layers that combine the IoT, SDN, NFV, and ML (mainly FL) components:

\textit{Application Layer:} Contains the applications that process and manage the data to be useful to the end user, as it is the front-end of the whole system. Through the service applications, the user can request services, which are provided using the IoT applications, SDN applications, and VNFs (by making queries intelligible to underlying layers).

\textit{Global Knowledge and Management Layer:} Centralizes network management and control through the head root SDN controller. This controller also acts as the FL aggregator, providing global intelligence by generating and broadcasting the global model to the multiple SDN controllers, and ensuring coordination with the IoT controller and NFV hypervisor.

\textit{Control Layer:} Contains the controllers of each domain (IoT, SDN, and NFV). Each oversees the underlying devices. SDN controllers contribute to the global intelligence by acting as FL learning nodes.

\textit{NFV Management and Orchestration (MANO) Layer:} Contains the management components of the network functions virtualization system that manages all of the virtualization processes of NFV.

\textit{Physical Layer:} Contains the physical resources of the system, including IoT sensing devices, network forwarding devices, and NFV environment infrastructure.

\section{Conclusion}
It is irrefutable that there is a critical need to provide agile and scalable IoT networks. In this paper, we have proposed a distributed intelligent network softwarization architecture for IoT networks. It takes advantage of SDN (in a distributed hierarchical design), NFV, and FL (as a distributed ML approach) to address IoT constraints and challenges.

As future work, we plan to implement and evaluate this architecture with consideration to Quality of Service (QoS) and IoT-specific metrics (e.g., energy efficiency). We plan to consider Deep Reinforcement Learning (DRL) as the learning approach on which the FL will be grounded, as it has shown remarkable results for decision problems in the networking environment. Then, our solution will be used to address IoT network challenges such as routing, load balancing, clustering, and VNF provisioning, while guaranteeing that the solution is the most optimal (by maximizing QoS indicators), and at the same time is resource aware and energy efficient.

\end{document}